\documentclass[conference]{IEEEtran}
\IEEEoverridecommandlockouts
\usepackage{cite}
\usepackage{amsmath,amssymb,amsfonts}
\usepackage{algorithmic}
\usepackage{graphicx}
\usepackage{textcomp}
\usepackage{xcolor}
\usepackage{subfigure}
\def\BibTeX{{\rm B\kern-.05em{\sc i\kern-.025em b}\kern-.08em
    T\kern-.1667em\lower.7ex\hbox{E}\kern-.125emX}}
    \usepackage{geometry}
\begin{document}
\title{{DAFT-Spread Affine Frequency Division Multiple Access for Downlink Transmission}}
\author{\IEEEauthorblockN{Yiwei~Tao\IEEEauthorrefmark{1},~Miaowen~Wen\IEEEauthorrefmark{1}, Yao~Ge\IEEEauthorrefmark{2}, Tianqi~Mao\IEEEauthorrefmark{3}, Lixia~Xiao\IEEEauthorrefmark{4}, and Jun~Li\IEEEauthorrefmark{5}
}
	\IEEEauthorblockA{\IEEEauthorrefmark{1}School of Electronic and Information Engineering, South China University of Technology, Guangzhou 510641, China}
	\IEEEauthorblockA{\IEEEauthorrefmark{2}Continental-NTU Corporate Lab, Nanyang Technological University, 639798, Singapore}
	\IEEEauthorblockA{\IEEEauthorrefmark{3}MIIT Key Laboratory of Complex-Field Intelligent Sensing, Beijing Institute of Technology, Beijing 100081, China}
	\IEEEauthorblockA{\IEEEauthorrefmark{4}School of Cyber Science and Engineering, Huazhong University of Science and Technology, Wuhan 430074, China}
	\IEEEauthorblockA{\IEEEauthorrefmark{5}School of Electronics and Communication Engineering, Guangzhou University, Guangzhou 510006, China}
		Email: eeyiweitao@mail.scut.edu.cn, eemwwen@scut.edu.cn, yao.ge@ntu.edu.sg,\\ maotq@bit.edu.cn,  lixiaxiao@hust.edu.cn, lijun52018@gzhu.edu.cn}
\vspace{+5mm}
\maketitle
\begin{abstract}
Affine frequency division multiplexing (AFDM) and orthogonal AFDM access (O-AFDMA) are promising techniques based on chirp signals, which are able to suppress the performance deterioration caused by Doppler shifts in high-mobility scenarios.
However, the high peak-to-average power ratio (PAPR) in AFDM or O-AFDMA is still a crucial problem, which severely limits their practical applications.
In this paper, we propose a discrete affine Fourier transform (DAFT)-spread AFDMA scheme based on the properties of the AFDM systems, named {\em DAFT-s-AFDMA} to significantly reduce the PAPR by resorting to the DAFT.
We formulate the transmitted time-domain signals of the proposed DAFT-s-AFDMA schemes with localized and interleaved chirp subcarrier allocation strategies.
Accordingly, we derive the guidelines for setting the DAFT parameters, revealing the insights of PAPR reduction.
Finally, simulation results of PAPR comparison in terms of the complementary cumulative distribution function (CCDF) show that the proposed DAFT-s-AFDMA schemes with localized and interleaved strategies can both attain better PAPR performances than the conventional O-AFDMA scheme.
\end{abstract}
\vspace{+1mm}
\section{Introduction}
With the rapid development of mobile communication technologies, next-generation communication networks expand the perspective to a wider range such as aeronautical communication, high-speed railway systems, vehicle-to-everything (V2X) systems, etc.~\cite{9689960,9861699,10152009}.
Under such high-mobility scenarios, the wireless channel exhibits fast time-varying properties with excessive Doppler shifts originated from the high-speed relative motion between the transceivers, which results in a series of challenges such as carrier frequency offset (CFO), inter-carrier interference (ICI), and channel estimation errors.
These factors seriously restrict the development of wireless communication services in high-mobility scenarios.
The current 4G/5G wireless communication techniques represented by orthogonal frequency division multiplexing (OFDM) are mainly designed for users with low/medium, and the high mobility greatly limits the coverage area and transmission reliability~\cite{9205980,6587554,5635467}.

Although fast time-varying fading will affect the transmission reliability of the communication systems, the short coherence time and fast variation of the fading channel also provide Doppler diversity that can be utilized to improve the system performance. In this context, a series of next-generation waveforms such as orthogonal chirp-division multiplexing (OCDM) and orthogonal time frequency space (OTFS) tailored for high-mobility communications have been developed~\cite{Ge_2021,10146020,9349154,10159363}. Affine frequency division multiplexing (AFDM) is another promising representative multicarrier modulation technique based on the discrete affine Fourier transform (DAFT)~\cite{9473655,9562168,10087310}.
AFDM can convert a linear time-varying (LTV) channel into a sparse quasi-static channel by adjusting the parameters of the DAFT according to the channel statistical distribution.
The channel taps caused by delay and Doppler shifts are fully separated in the discrete affine Fourier (DAF) domain, and the receiver can achieve full delay-Doppler diversity with advanced signal processing techniques.

Despite the considerable potential and significant attention garnered by AFDM for achieving reliable communications in high-mobility scenarios, the current explorations of its performance optimization and practical applications are still at their infancies.
To ensure accurate signal recovery at the receiver, some works have investigated channel estimation methods for AFDM schemes~\cite{9880774,zheng2024channel}.
To boost the transmission efficiency, the authors in~\cite{tao2023affine},~\cite{PrIM2024} and~\cite{10342712} attempted to incorporate index modulation (IM)~\cite{Wen2021IM} into the AFDM system recently.
Moreover, thanks to the high resolution for delay and Doppler, the AFDM was also used for integrated sensing and communication scenarios~\cite{10439996}.
\begin{figure*}[t]
	\center
	\includegraphics[width=7.2in,height=3.5in]{{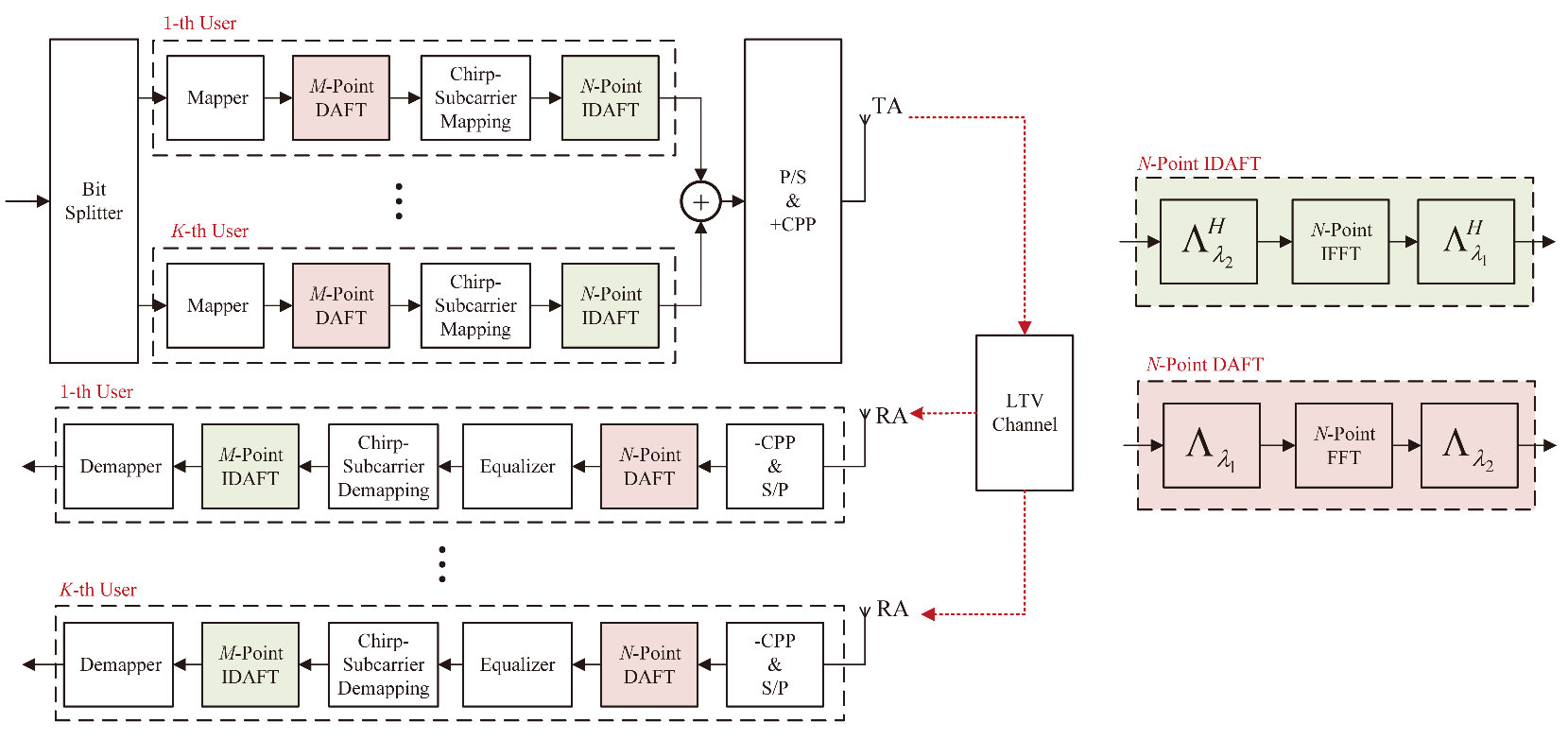}}
	\vspace{-0.4cm}
	\caption{{Transceiver structure of the proposed DAFT-s-AFDMA scheme.}}
	\label{system-model}  
	\vspace{-2mm}
\end{figure*}
To the best of our knowledge, there merely exists preliminary researches regarding the design of a multiple access strategy based on AFDM systems, and no literature has addressed this issue.
Inspired by the orthogonal frequency division multiple access (OFDMA) technique~\cite{5749771}, an effective approach is to divide the chirp subcarriers of AFDM to each user by orthogonal allocation.
However, as a multicarrier modulation technique, AFDM has a high peak-to-average power ratio (PAPR) as OFDM.
The high PAPR can not only result in non-linear distortion within the signal but also necessitates a more powerful power amplifier, which restricts the potential application scenarios for the orthogonal affine frequency division multiple access (O-AFDMA) system. On the other hand, as an alterative to OFDMA, the single carrier frequency division multiple access (SC-FDMA) has a low PAPR~\cite{5749771}.

Against the background,
we propose a DAFT-spread affine frequency division multiple access, named as {\em DAFT-s-AFDMA}.
Specifically, we use the $M$-point DAFT to preprocess the signal of each user in the DAF domain, and then use the $N$-point inverse DAFT (I-DAFT) to transform the signal into the time domain.
We derive the time-domain transmitted signal expressions for the proposed DAFT-s-AFDM schemes with localized and interleaved subcarrier allocation strategies.
On this basis, we analyse the parameter settings of the DAFT for PAPRs reduction of the proposed DAFT-s-AFDMA system.
Furthermore, we compare the PAPR of the proposed DAFT-s-AFDMA and O-AFDMA schemes by using the complementary cumulative distribution function (CCDF).
Simulation results show that both the DAFT-s-AFDMA schemes with localized and interleaved strategies have lower PAPR than O-AFDMA, and the DAFT-s-AFDMA scheme with interleaved strategies can achieve the same bit error rate (BER) performance as the O-AFDMA system.

The remainder of this paper is organized as follows. Section II describes the system model of the proposed DAFT-s-AFDMA scheme. Section III analyzes the time-domain signal expressions and DAFT parameter settings of the proposed scheme for PAPR reduction. Section IV discusses and analyzes the numerical results to evaluate the performance of the proposed scheme. Finally, conclusions are drawn in Section~V.

$Notations$: $(\cdot)^{T}$ and $(\cdot)^{H}$ represent the transpose and Hermitian transpose, respectively.
$[\cdot]_{N}$ and $\left\| {\cdot} \right\|$ represent modulo $N$ and Euclidean norm, respectively.
$\rm diag(\cdot)$ transforms a vector into a diagonal matrix. $\mathbb{C}^{M \times N}$ and $\mathbb{C}^{M\times1}$ are an $M\times N$ matrix and an $M\times1$ column vector with complex entries, respectively. ${\bf{I}}_{N}$ and ${\bf{0}}_{N}$ are an $N\times N$ identity matrix and an $N\times N$ zeros matrix, respectively.
${\mathcal{CN}}(0,{\sigma}^2)$ represents the complex Gaussian distribution with zero mean and variance of ${\sigma}^2$.
$\delta(\cdot)$ is the dirac delta function. $\otimes$ represents the convolution operator.
\vspace{+0mm}
\section{System Model of DAFT-s-AFDMA}
In this section, we describe the system model of the proposed DAFT-s-AFDMA scheme, where the corresponding transceiver structure is shown in Fig.~\ref{system-model}. We consider that either the transmitter or receiver is in motion with high relative velocities. Hence, the channel is modeled as an LTV channel.
\subsection{Transmitter}
As shown in Fig.~\ref{system-model}, we consider an AFDM symbol consisting of $N$ chirp subcarriers, and each user occupies $M=N/K$ chirp subcarriers, where $K$ is the total number of users.
After the constellation mapper, the transmission symbol vector for the $k$-th user is obtained as ${\bf x}_{k}\in{\mathbb{C}^{M\times1}}$, $k=0,\ldots,K-1$.
Then, an $M$-point DAFT is performed on ${\bf x}_{k}$ to obtain the symbol vector ${\bf x}'_{k}\in{\mathbb{C}^{M\times1}}$, which is given by
\begin{equation}\label{2.1}
{\bf x}'_{k}=\underbrace{{\bf\Lambda}_{{{\lambda}'_{2}}}{{{\bf F}_{M}}}{\bf\Lambda}_{{{\lambda}'_{1}}}}_{ M{\text{-point~DAFT}}}{\bf x}_{k},
\end{equation}
where ${\bf\Lambda}_{{{\lambda'_{1}}}}\!=\!{\rm diag}([e^{-j2\pi{\lambda'_1}{{0}^2}},\ldots,e^{-j2\pi {\lambda'_1}{{(M-1)}^2}}]^T)\in{\mathbb{C}^{M\times M}}$, ${\bf\Lambda}_{{{\lambda'_{2}}}}\!=\!{\rm diag}([e^{-j2\pi{\lambda'_2}{{0}^2}},\ldots,e^{-j2\pi {\lambda'_2}{{(M-1)}^2}}]^T)\in{\mathbb{C}^{M\times M}}$, ${\lambda}'_{1}$ and ${\lambda}'_{2}$ are two adjustable parameters of the $M$-point DAFT (we will discuss how to set their values to reduce the PAPR of the proposed DAFT-s-AFDMA system in the next section), and ${{\bf F}_{M}\in{\mathbb{C}}^{M\times M}}$ is the $M$-point normalized discrete fourier transform (DFT) matrix with ${\bf F}[\bar{m},{m}]=\frac{1}{\sqrt{M}}{e^{-j2\pi {\bar{m}}{m}/M}}$, and $m,{\bar m}\in\{0,1,\ldots,M-1\}$.

Next, we map the transmitted symbol vector ${\bf x}'_{k}$ to the chirp subcarriers of each user.
Define ${\bf\Gamma}_{k}\in{\mathbb{C}^{N\times M}}$ to be the chirp subcarrier mapping matrix of the $k$-th user. Therefore, the transmitted symbol ${\bf x}''_{k}\in{\mathbb{C}^{N\times1}}$ of the $k$-th user in the DAF domain is given by
\begin{equation}\label{2.2}
{{\bf x}''_{k}}={\bf\Gamma}_{k}{\bf x}'_{k}.
\end{equation}
Due to the orthogonal allocation, the chirp subcarrier mapping matrices of any two users are orthogonal, i.e., ${{\bf\Gamma}^{H}_{k}}{{\bf\Gamma}_{k'}}={\bf{0}}_{M}, k\ne k'$.

After the chirp-subcarrier mapping, the $N$-point IDAFT is performed on ${\bf x}''_{k}$ to generate the time-domain transmitted signal ${\bf s}_{k}$, and the superimposed transmitted signal of the $K$ users can be expressed as
\begin{equation}\label{2.3}
{\bf s}=\sum\limits_{k=0}^{K-1}{\underbrace{{\bf\Lambda}^{H}_{{{\lambda}_{1}}}{{{\bf F}_{N}^{H}}}{\bf\Lambda}^{H}_{{{\lambda}_{2}}}}_{ N{\text{-point~IDAFT}}}{\bf x}''_{k}},
\end{equation}
where ${\bf\Lambda}_{{{\lambda_{1}}}}\!=\!{\rm diag}([e^{-j2\pi{\lambda_1}{{0}^2}},\ldots,e^{-j2\pi {\lambda_1}{{(N-1)}^2}}]^{T})\in{\mathbb{C}^{N\times N}}$, ${\bf\Lambda}_{{{\lambda_{2}}}}\!=\!{\rm diag}([e^{-j2\pi{\lambda_2}{{0}^2}},\ldots,e^{-j2\pi {\lambda_2}{{(N-1)}^2}}]^{T})\in{\mathbb{C}^{N\times N}}$, and ${{\bf F}_{N}\in{\mathbb{C}}^{N\times N}}$ is the $N$-point normalized DFT matrix with ${\bf F}[\bar{u},{u}]=\frac{1}{\sqrt{N}}{e^{-j2\pi {\bar{u}}{u}/N}}$, and $u,{\bar u}\in\{0,1,\ldots,N-1\}$.
Finally, to avoid inter-block interference (IBI), a chirp periodic prefix (CPP) needs to be added to the transmitted signal, and the length of the CPP is greater than the maximum channel delay spread.

\subsection{Channel Model}
Consider an LTV channel with $P\ge1$ paths, and the channel impulse response (CIR) between the transmitter and the ${k}$-$\rm{th}$ user can be written as
\begin{equation}\label{AFDM.3}
h_{k}(\tau ,\nu )=\sum\limits_{\ell=1}^{P}{{{h}_{{k},\ell}}\delta (\tau -{{\tau }_{{k},\ell}})}{{e}^{j2\pi {{\nu }_{{k},\ell}}t}},
\end{equation}
where ${h}_{{k},\ell}\sim\mathcal{CN}(0,1/P)$, ${\tau}_{{k},\ell}\in[0,{\tau}_{\rm max}]$, and ${\nu}_{{k},\ell}\in[-{\nu}_{\rm max},{\nu}_{\rm max}]$ represent the CIR coefficient, actual delay spread, and actual Doppler spread index associated with the $\ell$-$\rm th$ path between the transmitter and the ${k}$-$\rm{th}$ user, respectively.
The normalized delay and Doppler shift of the $\ell$-$\rm th$ path are given by ${l}_{{k},\ell}={\tau}_{{k},\ell}{\Delta f}\in[0,{l}_{\rm{max}}]$ and ${\alpha_{{k},\ell}}={\nu}_{{k},\ell}{NT}$, respectively, where ${\Delta f}$ is the chirp subcarrier spacing and $T{\Delta f}=1$.
We define ${\alpha }_{{k},\ell}={\alpha }_{{\rm int},{k},\ell}+{\alpha }_{{\rm fra},{k},\ell}$ with ${\alpha }_{{\rm int},{k},\ell}\in[-{\alpha }_{\rm{max}},{\alpha }_{\rm{max}}]$ and $-\frac{1}{2}<{\alpha }_{{\rm fra},{k},\ell}<\frac{1}{2}$, which refer to the integer and fractional normalized Doppler shifts, respectively.
It is worth mentioning that this channel model is a generalized model that allows each delay tap to have different Doppler shift values, i.e., ${\ell}_{1},{\ell}_{2}=1,\ldots,P$, ${\tau}_{{k},{\ell}_{1}}={\tau}_{{k},{\ell}_{2}}$, and ${\alpha}_{{k},{\ell}_{1}}\ne{\alpha}_{{k},{\ell}_{2}}$. For the integer Doppler case, we set ${\lambda}_{1}=(2{\alpha}_{\rm max}+1)/2N$ and ${\lambda}_{2}$ to be an arbitrary irrational number to ensure that the system can achieve full delay and Doppler diversity~\cite{9473655,9562168,10087310}.
\subsection{Receiver}
Through the LTV channel and after removing the CPP, the time-domain received signal at the $k$-th user can be expressed as
\begin{equation}\label{2.5}
 \mathbf{r}_{k}={\mathbf{H}}_{k}{\mathbf s}+\mathbf{n},
\end{equation}
where ${\mathbf{H}}_{k}=\sum\nolimits_{\ell=1}^{P}{{h}_{{k},\ell}}{\mathbf{\Upsilon}_{{\text{Cpp}}_\ell}}{\mathbf\Delta_{{{\alpha }_{{k},\ell}}}}{\mathbf\Pi}^{l_{k,\ell}}$, ${\mathbf{\Upsilon}_{{\text{Cpp}}_\ell}}\in{\mathbb{C}^{N\times N}}$ is a diagonal matrix, ${\mathbf\Delta_{{{\alpha }_{{k},\ell}}}}={\rm diag}(e^{-j2\pi{{{\alpha }_{{k},\ell}}}u}, u=0,\ldots,N-1)$, ${\mathbf\Pi}$ is the $N\times N$ shift matrix~\cite{10087310}, and ${\bf n}\sim{\mathcal{CN}}({\bf 0},N_0{\bf{I}}_{N})\in{\mathbb{C}^{N\times1}}$ is the noise vector.
Next, an $N$-point DAFT is performed on the received signal $\mathbf{r}_{k}$ to recover the DAF-domain signal of the $k$-th user, which is given by
\begin{align}\label{2.6}
   {{\mathbf{y}}_{k}}&={{\mathbf{\Lambda }}_{{{\lambda }_{2}}}}{{\mathbf{F}}_{N}}{{\mathbf{\Lambda }}_{{{\lambda }_{1}}}}({{\mathbf{H}}_{k}}\mathbf{s}+\mathbf{n}), \nonumber\\
 & =\underbrace{{{\mathbf{\Lambda }}_{{{\lambda }_{2}}}}{{\mathbf{F}}_{N}}{{\mathbf{\Lambda }}_{{{\lambda }_{1}}}}{{\mathbf{H}}_{k}}\mathbf{\Lambda }_{{{\lambda }_{1}}}^{H}\mathbf{F}_{N}^{H}\mathbf{\Lambda }_{{{\lambda }_{2}}}^{H}}_{{\bf H}_{\rm eff}}\sum\limits_{k=0}^{K-1}{{\bf x}''_{k}}+{{\mathbf{\Lambda }}_{{{\lambda }_{2}}}}{{\mathbf{F}}_{N}}{{\mathbf{\Lambda }}_{{{\lambda }_{1}}}}\mathbf{{n}}, \nonumber\\
 & ={{\mathbf{H}}_{{\rm eff},k}}\sum\limits_{k=0}^{K-1}{{\bf x}''_{k}}+\mathbf{\tilde{n}}.
\end{align}

Due to the high complexity of maximum likelihood (ML) detection, we apply the linear minimum mean square error (MMSE) equalization to eliminate the interference, yielding
\begin{align}\label{2.7}
{{\widehat{\bf x}''}_{{\rm MMSE}}} &={{\bf H}}_{{\rm eff},k}^{H}{{\left( {{{\bf H}}_{{\rm eff},k}}{{\bf H}}_{{\rm eff},k}^{H}+{\frac{1}{\gamma}}{{\bf I}_{N}} \right)}^{-1}}{\bf y}_{k}\nonumber\\
 & =\sum\limits_{k=0}^{K-1}{{\bf x}''_{k}}+{\mathbf{\vartheta }},
\end{align}
where ${\gamma}={E}_{s}/{N}_{0}$ with ${E}_{s}$ representing the average energy per transmitted symbol, ${{\widehat{\bf x}''}_{{\rm MMSE}}}$ is the estimated symbol vector, and ${\mathbf{\vartheta}}\in{{\mathbb{C}}^{N\times 1}}$ represents the estimated error vector. Afterwards, by using the subcarrier mapping matrix of the $k$-th user and the $M$-point IDAFT matrix, we can obtain the transmitted symbols of the $k$-th user as follows
\begin{align}\label{2.8}
{{\widehat{\bf x}_{k}}} &={\mathbf A}^{H}{\bf\Gamma}^{H}_{k}\sum\limits_{k=0}^{K-1}{{\bf x}''_{k}}+{\mathbf A}^{H}{\bf\Gamma}^{H}_{k}{\mathbf{\vartheta }}\nonumber\\
 & ={\mathbf x}_{k}+\sum\limits_{k'=0,k'\ne k}^{K-1}{\mathbf A}^{H}\underbrace{{{\bf\Gamma}^{H}_{k}{\bf\Gamma}_{k}}}_{={\mathbf{0}_{M}}}{\mathbf A}{\bf x}_{k'}+{\mathbf A}^{H}{\bf\Gamma}^{H}_{k}{\mathbf{\vartheta }},
\end{align}
where ${\mathbf A}={{\bf\Lambda}_{{{\lambda}'_{2}}}{{{\bf F}_{M}}}{\bf\Lambda}_{{{\lambda}'_{1}}}}$.
\vspace{+3mm}
\section{PARR Analysis For DAFT-s-AFDMA}\label{section.3}
\begin{figure}[t]
	\center
	\includegraphics[width=3.6in,height=1.9in]{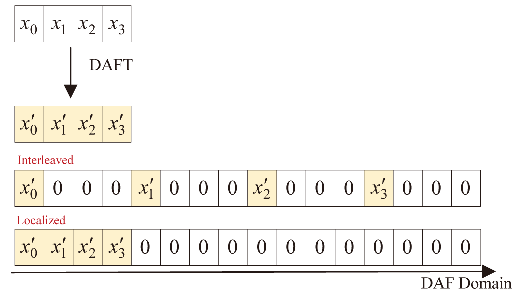}
	\vspace{-0.4cm}
	\caption{{An example of the proposed DAFT-s-AFDMA schemes with the transmitted symbols in the DAF domain for the $k$-th user, where $N=16$, $M=4$ and $K=4$.}}
	\label{strategies}  
	\vspace{-2mm}
\end{figure}

In this section, we analyze the PAPR of the proposed DAFT-s-AFDMA scheme. As shown in Fig.~\ref{strategies}, we consider two chirp-subcarrier allocation strategies, i.e., the interleaved and localized strategies.
The time-domain transmitted signal expressions in this two strategies are further analyzed to derive the setting of the DAFT parameters that can reduce the PAPR.

The PAPR of the time-domain transmitted signal of the DAFT-s-AFDMA scheme after discrete sampling is defined~as
\begin{align}\label{3.1}
{\rm PAPR}=\frac{{P}_{\rm max}}{{P}_{\rm avg}}=\frac{\underset{u=0,\ldots ,N-1}{\mathop{\text{max}}}\{\left\|s[u]\right\|^{2}\}}
{\frac{1}{N}\sum\limits_{u=0}^{N-1}{\left\|s[u]\right\|^{2}}},
\end{align}
where $s[u]$ is the $u$-th element of the time-domain transmitted symbol vector $\bf s$.
To quantify the PAPR performance of the proposed DAFT-s-AFDMA scheme, we can calculate the probability that the system PAPR exceeds a certain level ${\rm PAPR_0}$, i.e., ${\rm PAPR>PAPR_0}$.
Therefore, we use the CCDF to evaluate the PAPR performance of the system, which can be formulated as
\begin{align}\label{3.2}
{\rm CCDF}={\rm Pr}({\rm PAPR>PAPR_0}).
\end{align}
\begin{figure}[t]
	\center
	\includegraphics[width=3.6in,height=1.9in]{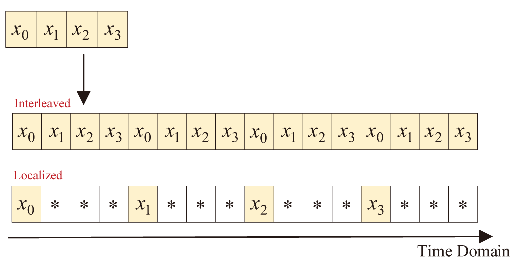}
	\vspace{-0.4cm}
	\caption{{An example of the proposed DAFT-s-AFDMA schemes with the transmitted symbols in the time domain for the $k$-th user, where $N=16$, $M=4$ and $K=4$.}}
	\label{strategies2}  
	\vspace{-4mm}
\end{figure}
\subsection{Chirp Subcarrier Mapping with Interleaved Strategy}
For the proposed DAFT-s-AFDMA scheme with the interleaved strategy, the transmitted signal of the $k$-th user in the DAF domain can be written as
\begin{equation}\label{3.3}
{{x}''_{k}[u]}=\left\{\begin{array}{ll}
	{{x}'_{k}}\left[\frac{u}{K}\right], & {u}=K{m}, \\
	0, & {\rm otherwise, }
\end{array}\right.
\end{equation}
where ${{x}'_{k}}[m]\!\!=\frac{1}{\sqrt{M}}\sum\nolimits_{\bar{m}=0}^{M-1}{x[\bar m]}{{e}^{-j2\pi \left( {{{{\lambda }'}}_{1}}{{m}^{2}}+\frac{m\bar{m}}{M}+{{{{\lambda }'}}_{2}}{{{\bar{m}}}^{2}} \right)}}$ and $0\le\!{m}\le\!{M-1}$.
Let $u=Mq+{\bar{m}}$, where $0\le \bar m\le {M-1}$ and $0\le q\le {K-1}$.
Without loss of generality, we derive the $u$-th time-domain transmitted symbol by using the $N$-point IDAFT, as follows
\begin{align}\label{3.4}
   &{{s}_{k}}[u]=\frac{1}{\sqrt{N}}\sum\limits_{\bar{u}=0}^{N-1}{{{{{x}''}}_{k}}[\bar{u}]}{{e}^{j2\pi \left( {{\lambda }_{1}}{{u}^{2}}+\frac{u\bar{u}}{N}+{{\lambda }_{2}}{{{\bar{u}}}^{2}} \right)}} \nonumber\\
  =&\frac{1}{\sqrt{MK}}\sum\limits_{\bar{m}=0}^{M-1}{{{{{x}'}}_{k}}[\bar{m}]}{{e}^{j2\pi \left( {{\lambda }_{1}}{{(Mq+\bar{m})}^{2}}+\frac{(Mq+m)K\bar{m}}{MK}+{{\lambda }_{2}}{{K}^{2}}{{{\bar{m}}}^{2}} \right)}} \nonumber\\
  =&\frac{1}{\sqrt{K}}\left( \frac{1}{\sqrt{M}}\sum\limits_{\bar{m}=0}^{M-1}{{{{{x}'}}_{k}}[\bar{m}]{{e}^{j2\pi \left( {{{{\lambda }}}_{1}}{{m}^{2}}+\frac{\bar{m}m}{M}+{{{{\lambda }'}}_{2}}{{{\bar{m}}}^{2}} \right)}}} \right) \nonumber\\
 &\times{{e}^{j2\pi \left( {{\lambda }_{2}}{{K}^{2}}{{{\bar{m}}}^{2}}-{{{{\lambda }'}}_{2}}{{{\bar{m}}}^{2}} \right)}}{{e}^{j2\pi \left( {{\lambda }_{1}} {{(Mq+m)}^{2}}-{{{{\lambda }'}}_{1}}{{m}^{2}} \right)}}.
\end{align}
Now, we can observe the relationship between the parameters ${{{{\lambda }'}}_{1}}$, ${{{{\lambda }'}}_{2}}$, ${{{{\lambda }}}_{1}}$, and ${{{{\lambda }}}_{2}}$ in the $M$-point DAFT and $N$-point IDAFT and determine the specific setting to reduce the PAPR. From above analysis, we set ${{{{\lambda }}}_{1}}={{{{\lambda }'}}_{1}}$ and ${{{{\lambda }}}_{2}}{K^2}={{{{\lambda' }}}_{2}}$, such that~\eqref{3.4} can be further expressed in the following form:
\begin{align}\label{3.5}
   {{s}_{k}}[Mq+{\bar{m}}]
  =&\frac{1}{\sqrt{K}}x_{k}[\bar m]{{e}^{j2\pi \left( {{\lambda }_{1}}{{((Mq)^{2}+2Mmq)}} \right)}}.
\end{align}
Hence, the time-domain transmitted signal ${\bf s}_{k}$ can be reduced to a period-repeated version of the original input signal ${\bf x}_{k}$, as shown in Fig.~\ref{strategies2}. With this parameter setting, the PAPR of the proposed DAFT-s-AFDMA scheme with the interleaved chirp-subcarrier allocation is similar to that of the conventional single-carrier scheme.

\subsection{Chirp Subcarrier Mapping with Localized Strategy}
Similarly, for the proposed DAFT-s-AFDMA scheme with the localized strategy, the $k$-th user transmitted signal in the DAF domain can be expressed as
\begin{equation}\label{3.6}
{{x}''_{k}[u]}=\left\{\begin{array}{ll}
	{{x}'_{k}}\left[m\right], & 0\le u \le{M-1}, \\
	0, & M\le u \le{N-1}.
\end{array}\right.
\end{equation}
We define $u=K{\bar m}+q$, and the $u$-th time-domain transmitted symbol of the $k$-th user can be further calculated as
\begin{align}\label{3.7}
  &{{s}_{k}}[u]=\frac{1}{\sqrt{N}}\sum\limits_{\bar{u}=0}^{N-1}{{{{{x}''}}_{k}}[\bar{u}]}{{e}^{j2\pi \left( {{\lambda }_{1}}{{u}^{2}}+\frac{u\bar{u}}{N}+{{\lambda }_{2}}{{{\bar{u}}}^{2}} \right)}} \nonumber\\
 &\!\!=\frac{1}{\sqrt{KM}}\!\!\sum\limits_{{m}=0}^{M-1}\!\!{{{{{x}'}}_{k}}[m]}{{e}^{j2\pi \left( {{\lambda }_{1}}{{(K\bar{m}+q)}^{2}}+\frac{(K\bar{m}+q){m}}{MK}+{{\lambda }_{2}}{{{m}}^{2}} \right)}}.
\end{align}
If $q=0$, ${{{{\lambda }}}_{1}}={{{{\lambda }'}}_{1}}$ and ${{{{\lambda }}}_{2}}={{{{\lambda' }}}_{2}}$,~\eqref{3.7} can be simplified as
\begin{align}\label{3.8}
  {{s}_{k}}[K{\bar m}]&=\frac{1}{\sqrt{K}}x_{k}[\bar m]e^{j2\pi((K\bar{m})^2-\bar{m}^2)}.
\end{align}
When $q\ne0$, the transmitted signal at these locations is the sum of the original input signal ${\bf x}_{k}$ with different complex weighting values, which increases the PAPR~\cite{5749771}.

In the next section, we will test the PAPR performance defined in~\eqref{3.1} and~\eqref{3.2} for both the interleaved and localized chirp-subcarrier allocations.

\vspace{+3mm}
\section{Simulation Results and Discussions}
In this section, we first evaluate the BER performance of the proposed DAFT-s-AFDMA schemes with localized and interleaved strategies, and O-AFDMA scheme by using the MMSE detection.
Then, we numerically compare the PAPRs of the proposed DAFT-s-AFDMA with two chirp subcarrier allocation strategies, and O-AFDMA scheme by using the CCDF.
Without loss of generality, we set the carrier frequency as $4~{\rm{GHz}}$, the subsymbol spacing in the DAF domain as $15~{\rm{kHz}}$, the number of chirp subcarriers as $N=1024$, and the modulation alphabet as QPSK.
We use the Jakes Doppler spectrum approach to generate the Doppler shifts, i.e., $\alpha_{\epsilon,\ell}={\alpha_{\rm max}{\rm{cos}}(\theta_{\epsilon,\ell})}$, where $\theta_{\epsilon,\ell}\in[-\pi,\pi]$ obeys a uniform distribution~\cite{9293173}.
In addition, we set
${\lambda}_{1}={\lambda}'_{1}=(2{\alpha}_{\rm max}+1)/2N$, ${\lambda}_{2}K^2={\lambda}'_{2}=\pi$ for the interleaved strategy, and ${\lambda}_{2}={\lambda}'_{2}=\pi$ for the localized strategy to reduce the PAPR of the proposed DAFT-s-AFDMA system. The signal-to-noise ratio (SNR) is defined as ${E}_{b}/{N}_{0}$, where ${E}_{b}$ is the bit~energy.

\begin{figure}[t]
	\center
	\includegraphics[width=3.5in,height=2.7in]{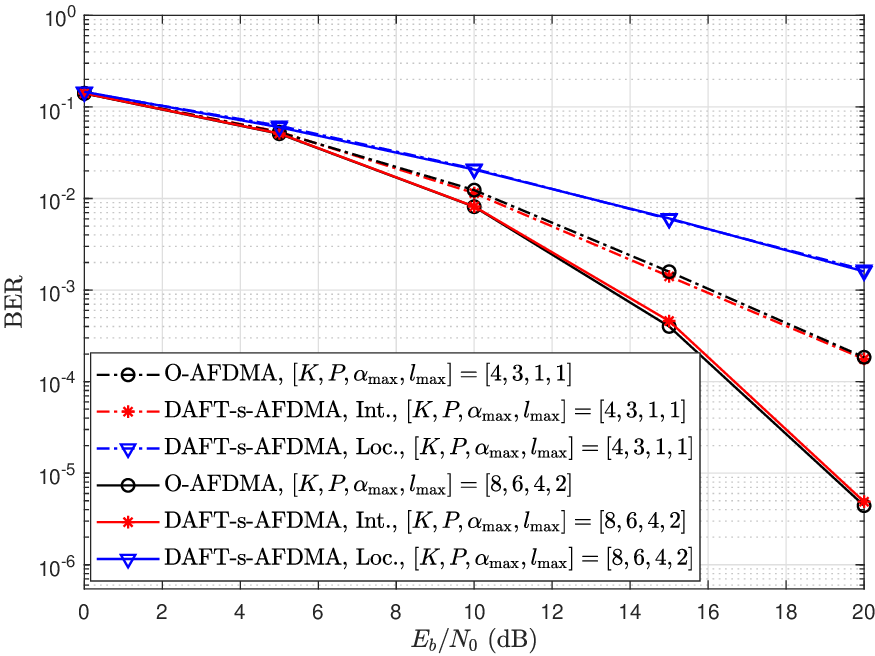}
	\vspace{-0.1cm}
	\caption{{Comparison of the BER performance of the proposed DAFT-s-AFDMA schemes with interleaved and localized strategies, and the conventional O-AFDMA scheme, where $N=1024$ and $[K, P,{\alpha_{\rm max}},{l_{\rm max}}]=[4, 3, 1, 1]$ or [8, 6, 4, 2]. }}
	\label{BER_V1}  
	\vspace{-3mm}
\end{figure}
\begin{figure}[t]
	\center
	\includegraphics[width=3.5in,height=2.7in]{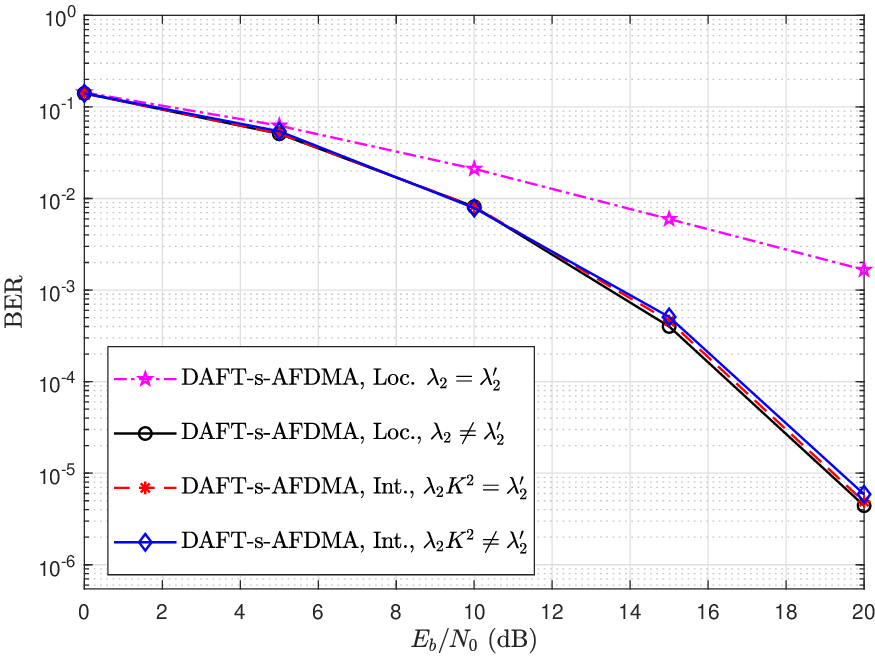}
	\vspace{-0.1cm}
	\caption{{The effect of DAFT parameter $\lambda_2$ and $\lambda'_2$ on the BER performance of the proposed DAFT-s-AFDMA schemes with interleaved and localized strategies, where $N=1024$, and $[K, P,{\alpha_{\rm max}},{l_{\rm max}}]$=[8, 6, 4, 2]. }}
	\label{BER_V2}  
	\vspace{-1mm}
\end{figure}

Figure~\ref{BER_V1} illustrates the BER performance of the proposed DAFT-s-AFDMA schemes and O-AFDMA scheme over the LTV channnel.
We set two different sets of parameters for the simulation, i.e., $[K, P, \alpha_{\rm max}, l_{\rm max}]=[4,3,1,1]$ and $[K, P, \alpha_{\rm max}, l_{\rm max}]=[8,6,4,2]$. One can notice that the proposed DAFT-s-AFDMA scheme with the interleaved strategy achieves similar performance to the O-AFDMA scheme, while the BER performance of the DAFT-s-AFDM scheme with the localized strategy deteriorates.
Moreover, one can observe that the performance of both the proposed DAFT-s-AFDMA scheme with the interleaved strategy and O-AFDMA scheme improves significantly as the number of channel paths $P$ increases.
On the contrary, the performance of the DAFT-s-AFDMA scheme with the localized strategy is consistent for the parameter settings of $P=3$ and $P=6$. This shows that the localized strategy leads to a decrease in the diversity order of the system.

In Fig.~\ref{BER_V2}, we investigate the impact of the DAFT parameter on the BER performance of the proposed DAFT-s-AFDM systems with localized and interleaved strategies.
Since the setting of $\lambda_1$ is required to achieve the full channel diversity~\cite{10087310}, we only discuss the setting of $\lambda'_2$ and $\lambda_2$. One can observe form Fig.~\ref{BER_V2} that the BER performance of the  interleaved DAFT-s-AFDM scheme is not affected by the $\lambda'_2$ and $\lambda_2$ setting, whereas the BER performance of the localized DAFT-s-AFDMA scheme deteriorates when using the $\lambda'_2$ and $\lambda_2$ setting for low PAPR.

\begin{figure}[htp]
	\center
	\includegraphics[width=3.5in,height=2.7in]{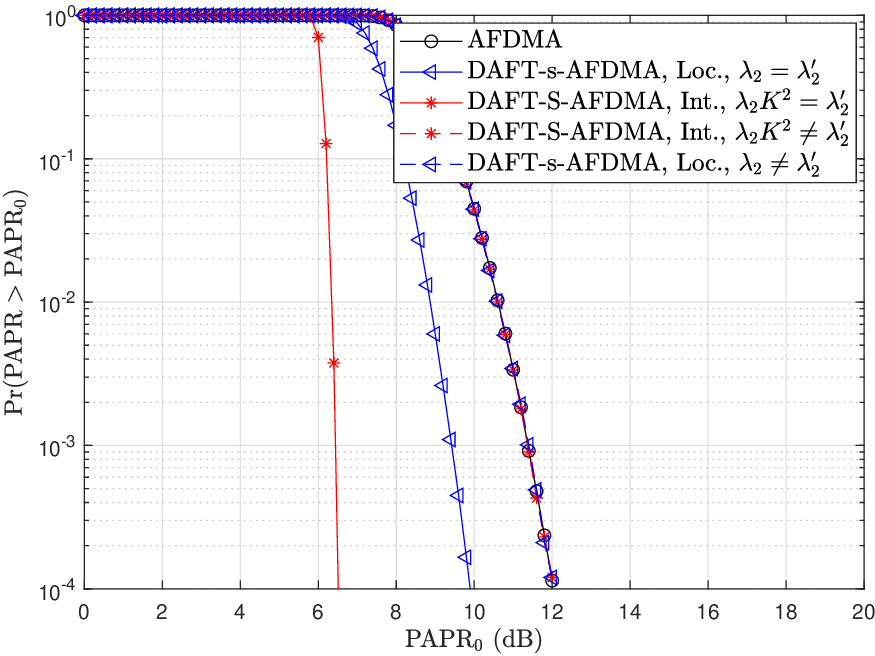}
	\vspace{-0.2cm}
	\caption{{Comparison of the CCDF of PAPRs of the proposed DAFT-s-AFDMA schemes with interleaved and localized strategies, and the conventional O-AFDMA scheme, where $N=1024$, and $K=4$.}}
	\label{1024_U4}  
	\vspace{-2mm}
\end{figure}

In Fig.~\ref{1024_U4}, we further compare the PAPRs of the proposed DAFT-s-AFDMA schemes with interleaved and localized strategies, and the conventional O-AFDMA scheme.
One can observe that the proposed DAFT-s-AFDMA schemes with both the localized and interleaved strategies achieve better PAPR performance than the O-AFDMA scheme with the DAFT parameter settings in Section III. For instance, when targeting the CCDF at a level of $10^{-4}$, the proposed DAFT-s-AFDMA schemes with the localized and interleaved strategies yield about $5.5~{\rm dB}$ and $2~{\rm dB}$ gains with respect to the O-AFDMA scheme, respectively. However, when the derived DAFT parameter setting guidelines are not satisfied, the proposed DAFT-s-AFDM scheme fails to reduce the PAPR. Moreover, the interleaved strategy has lower PAPR than the localized strategy, which is consistent with the analysis in Section~\ref{section.3} as the localized strategy can only partially tackle the PAPR.

\vspace{-2mm}
\section{Conclusion}
In this paper, we proposed a DAFT-s-AFDMA scheme for downlink multiple access, where the additional DAFT operation is applied to reduce the PAPR of the system.
We investigated the effect of two chirp subcarrier allocation strategies, i.e., localized and interleaved, on the PAPR of the proposed DAFT-s-AFDMA system, and derived the time-domain transmitted signal expression.
On this basis, we analyzed the setting of DAFT parameters to effectively reduce the PAPR of the proposed DAFT-s-AFDMA system.
Simulation results demonstrated that the proposed DAFT-s-AFDMA schemes with localized and interleaved strategies achieve better PAPR performance than the O-AFDMA scheme.
Moreover, the proposed DAFT-s-AFDMA scheme with interleaved strategy achieves similar BER performance to the O-AFDMA scheme, but the DAFT-s-AFDMA scheme with localized strategy will cause performance degradation due to a diversity loss.
Therefore, the interleaved DAFT-s-AFDMA scheme in the downlink can reduce the system PAPR without BER performance loss.
In the future, we will investigate low-complexity detection algorithms and uplink multiple-access schemes based on the AFDM architecture.

\bibliographystyle{IEEEtran}  
\bibliography{IEEEabrv,ref}

\end{document}